\documentclass[aip,apl,reprint]{revtex4-1}

\usepackage{amsmath}
\usepackage{natbib}
\usepackage{epsfig,graphicx}
\usepackage{bm}

\begin{document}

\title{Transparent ion trap with integrated photodetector}

\author{Amira M. Eltony} \email[]{aeltony@mit.edu}
\author{Shannon X. Wang}
\author{Gleb M. Akselrod}
\author{Peter F. Herskind}
\author{Isaac L. Chuang}
\affiliation{Center for Ultracold Atoms, Research Laboratory of Electronics and Department of Physics, Massachusetts Institute of Technology, Cambridge, Massachusetts, 02139, USA}

\date{\today}

\begin{abstract}
\noindent Fluorescence collection sets the efficiency of state detection and the rate of entanglement generation between remote trapped ion qubits. Despite efforts to improve light collection using various optical elements, solid angle capture is limited to $\approx10\%$ for implementations that are scalable to many ions. We present an approach based on fluorescence detection through a transparent trap using an integrated photodetector, combining collection efficiency approaching 50\% with scalability. We microfabricate transparent surface traps with indium tin oxide and verify stable trapping of single ions. The fluorescence from a cloud of ions is detected using a photodiode sandwiched with a transparent trap.
\end{abstract}

\maketitle

Integrated trapped ion quantum computation systems have recently made advances building on silicon chip technology by incorporating a variety of devices including optical fibers \cite{VanDevender2010, Kim2011c, Brady2011}, MEMs cantilevers \cite{Stick2005}, and control electronics. This integration seeks to improve gate fidelities, fault tolerance thresholds, and scalability, but state detection efficiency relies on fluorescence detection, which is largely still implemented with bulk optics and conventional photomultipliers or image-intensified charge coupled detectors. Scaling-up to dense arrays of trapped ions will require efficient light collection from many ions in parallel. For distributed architectures, with remote atomic nodes connected by photons, light collection is also a critical factor determining the efficiency of remote entanglement generation \cite{Luo2009}.

The conventional approach to light collection places a high numerical aperture objective near the ion, and detects light with a photomultiplier or charge-coupled detector located outside of the vacuum chamber, resulting in solid angle capture of less than 5\%. There are various proposals to enhance atom-photon coupling within a scalable architecture. An array of Fresnel lenses in place of a bulk lens can provide more efficient light collection from multiple ions \cite{Streed2011, Jechow2011}. Or, a micro mirror embedded into a planar trap can improve solid angle collection from an ion trapped above \cite{Herskind2011, Noek2010} to as much as $\approx10\%$ \cite{TrueMerrill2011}. Greater solid angle capture is possible, at the cost of scalability, by placing the trapping site at the focus of a spherical \cite{Shu2011} or parabolic \cite{Maiwald2012} mirror. Similarly, the ion can be trapped within a high-finesse optical cavity for enhanced light collection into a single mode \cite{Mundt2002, Keller2004}. Integration of a planar ion trap with an optical fiber has been demonstrated \cite{Kim2011c, Brady2011}, although the solid angle collection is low: $\approx3.5\%$ \cite{VanDevender2010}.

A different concept is to collect fluorescence \emph{through} a planar trap via the underutilized $2\pi$ solid angle below the ion. A surface electrode trap made of transparent materials would in principle allow for collection efficiency approaching 50\%. Combining such a transparent trap with an array of detectors beneath opens up the possibility of massively parallel light collection. Indium tin oxide (ITO), an electrical and optical conductor commonly used for applications such as touch screens and LCD displays, is a natural choice for transparent electrodes. But, trapping ions with ITO electrodes poses significant difficulty because ITO has a resistivity about 1000 times higher than metals typically used for trap electrodes, and ITO is an oxide, making laser-induced charging of its surface a concern \cite{Wang2011}.

Motivated by these challenges, we present results from transparent surface electrode ion traps, fabricated using ITO. Two ITO traps are tested at cryogenic temperatures (4~K and 77~K), where low pressure is achievable within a short time frame, and heating of the ion's motional state is suppressed \cite{Labaziewicz2008}. $^{88}$Sr$^+$ ions are trapped 100~$\mu$m above the trap surface, which is heat sunk to the 4~K stage of a bath cryostat. The fluorescence emitted on the Doppler cooling transition (5S$_{1/2}$ $\leftrightarrow$ 5P$_{1/2}$) at 422~nm is collected for state detection. We observe stable trapping in a first ITO trap. We then demonstrate a proof-of-principle prototype for scalable fluorescence detection in a second ITO trap by collecting light from a trapped ion cloud using a standard photodiode mounted below. Finally, we propose a highly-efficient and compact ``entanglement unit'' based on this design.

\begin{figure}[tb]
\includegraphics[width=3.3in]{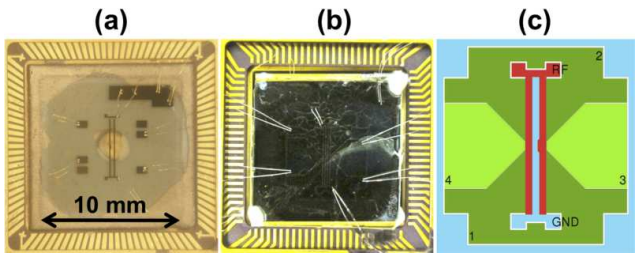}
\caption{\label{fig:traps} (a) ITO-4K trap mounted in a CPGA; 50~nm of Au is visible on the RF electrodes and contact pads for wire bonding. (b) ITO-PD trap mounted on photodiode in a CGPA; with only 5~nm of Au on the RF electrodes, the photodiode is visible through the trap. (c) Diagram of trap geometry showing RF electrodes (red), ground electrodes (blue), and DC electrodes (green).}
\end{figure}

The trap geometry used is a well established five-electrode design\cite{Labaziewicz2008} (see Figure \ref{fig:traps}(c)), with trap frequencies in the range of 0.8-1.3~MHz for an RF frequency of 35~MHz, and a trap depth of about 300~meV. Trap fabrication begins with optical lithography using NR9-3000PY photoresist on quartz substrate. Next, ITO is deposited by RF sputtering with argon gas at a rate of 0.5~\AA/s. Finally, the resist pattern is transferred to the ITO via lift-off with RD6. The resulting optical transmission of the ITO samples (including the polished quartz substrate), measured over a 4~mm$^2$ area with a 422~nm light source at room temperature, averages to $\approx60\%$, which is about 10\% lower than expected for commercial films \cite{Kim1999}. The measured resistivity of ITO varies from $1\times10^{-5}$~$\Omega$m to $2\times10^{-5}$~$\Omega$m for different sputtering runs. To improve the conductivity of the RF electrodes, an additional lithography step is performed to deposit a thin gold layer on the RF electrodes only.

The first ITO trap, ITO-4K (see Figure \ref{fig:traps}(a)), has 400~nm of ITO for all trap electrodes, and an additional 50~nm of gold on the RF electrodes only, bringing the resistivity down to $\approx 2\times10^{-8}$~$\Omega$m, which is comparable to a fully-metal trap. The second ITO trap, ITO-PD (see Figure \ref{fig:traps}(b)), also has 400~nm of ITO for all trap electrodes, but only 5~nm of gold on the RF electrodes for improved transparency ($\approx60\%$ for 5~nm of Au \cite{Stognij2003}) with a resistivity of $\approx 3\times10^{-8}$~$\Omega$m. This second trap is sandwiched with a photodetector. For a proof-of-concept demonstration, a commercially available PIN photodiode (Advanced Photonix PDB-C613-2) is used. The photodiode efficiency drops significantly at cryogenic temperatures due to carrier freeze-out. Measurements at 422~nm reveal that the photodiode responsivity changes little from $\approx0.1$~A/W as it is cooled from room temperature to 77~K, but plummets to $\approx0.01$~A/W at 4~K. For this reason, measurements with the photodiode are performed at 77~K.

For ITO-4K, the estimated trap-surface temperature was 6~K. Single $^{88}$Sr$^+$ ions were stably trapped with a lifetime of several hours, comparable to metal traps. No significant change in the micromotion amplitude was measured after crashing either a 405~nm or a 461~nm laser beam with intensity $\approx5$~mW/mm$^2$ into the center of the trap\cite{Wang2011} for 10~minutes, indicating that charging is not a major problem for ITO traps operated at cryogenic temperatures.

\begin{figure}[tb]
\includegraphics[width=3.3in]{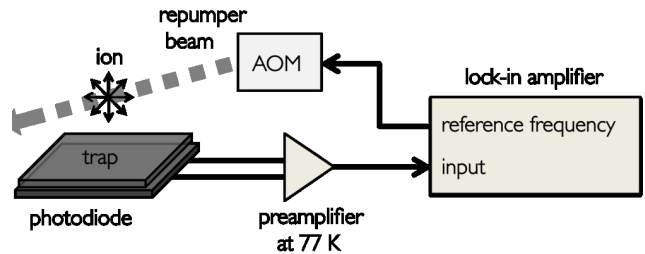}
\caption{\label{fig:diagram} Apparatus for detection of ion fluorescence through a transparent trap using a photodiode mounted below. The acousto-optic modulator (AOM) is used to modulate the repumper for lock-in detection.}
\end{figure}

ITO-PD has an estimated trap-surface temperature of 77~K. Because the photodiode used has no internal gain mechanism, the resulting picoamp-scale photocurrent is difficult to distinguish from electrical noise, making lock-in detection essential. The intensity of the laser addressing the 4D$_{3/2}$ $\leftrightarrow$ 5P$_{3/2}$ transition is chopped at 300~Hz, resulting in modulation of the ion fluorescence as population is successively trapped then pumped from the metastable 4D$_{3/2}$ state during Doppler cooling. Inside the cryostat, a custom preamplifier circuit mounted to the 77~K shield amplifies the signal with low added noise before it is input to a lock-in amplifier (Stanford Research Systems SR530) outside the chamber, as shown in Figure \ref{fig:diagram}. The expected signal for $\approx50$~ions in this setup is estimated in Table \ref{tab:estimate} and compared with our conventional bulk optics and photomultiplier (PMT) setup.

\begin{table}[tb]
\begin{tabular}{cccccc}
\hline\hline
 & Light & Power & Detector & Photodiode & Lock-in \\
 & collection & at the & quantum & current & amplifier \\
 & efficiency & detector & efficiency & & output \\
\hline
ITO-PD & 30\% & 60 pW & 30\% & 6 pA & 120 mV \\
PMT & 5\% & 10 pW & 20\% & n.a. & n.a. \\
\hline\hline
\end{tabular}
\caption{\label{tab:estimate} Comparison of photodiode and photomultiplier (PMT) collection efficiencies, with expected signal values for 50~ions, assuming a scattering rate of $\approx10^7$~photons/s per ion at 422~nm, resulting in $\approx200$~pW of total fluorescence into $4\pi$~solid angle.}
\end{table}

\begin{figure}[tb]
\includegraphics[width=3.3in]{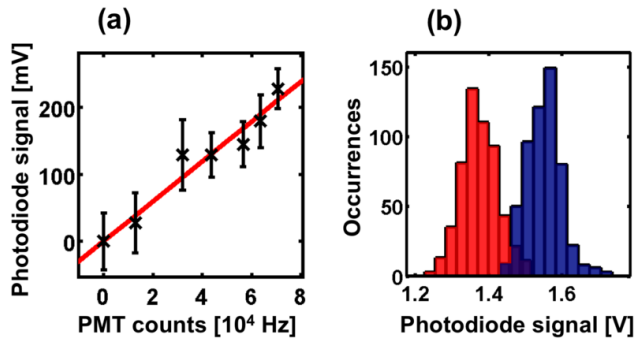}
\caption{\label{fig:data} (a) Photodiode voltage and photomultiplier count rate, both background subtracted, during loading of an ion cloud. Each point is averaged over 30~seconds. (b) Histogram of photodiode voltages over a period of several minutes without ions (red), and after loading an ion cloud (blue).}
\end{figure}

The pressure in the cryostat, when operated at 77~K, is insufficient ($\approx1\times10^{-7}$~Torr) for stable trapping, so a cloud of ions was continually loaded. Figure \ref{fig:data}(a) plots the photodiode voltage during initial loading of an ion cloud against the photon counts for light collected at the same time using bulk optics and a photomultiplier, indicating that the photodiode response is proportional to the fluorescence rate. Variation in the photodiode signal is likely due to scatter from the modulated repumper beam. Figure \ref{fig:data}(b) compares the photodiode voltage before and after an ion cloud was loaded, showing that the ion cloud fluorescence is distinguishable from the background. Over a measurement interval of a few minutes, the detected signal from the ion cloud averages to $175 \pm 49$~mV, which is consistent with the signal predicted above for $\approx50$~ions.

These experiments indicate that significant improvements in quantum state detection are possible using transparent traps with integrated detectors. Assuming a noiseless amplifier, our photodiode signal could be used to distinguish between quantum states with greater than 99\%~fidelity with a 1~ms integration time. More generally, nearly 50\%~solid-angle collection is possible by using a photodiode with a large active area or focusing fluorescence onto the photodiode using additional optics below the trap. Then, the only losses before the detector occur in the ITO film and in the substrate. For commercially available ITO films these losses could be as low at 10\% \cite{Kim1999}. Replacing the photodiode by a device with an internal gain mechanism such as the Visible Light Photon Counter (VLPC), which has a quantum efficiency of 88\% at 694~nm and 4~K \cite{McKay2009}, would allow a total detection efficiency of nearly 40\%, compared to the typical 1-5\% possible with a conventional photomultiplier and bulk optics. For our system, this would mean reducing the time required for quantum state detection with 99\% fidelity from the current 200~$\mu$s to only 5~$\mu$s.

\begin{figure}[tb]
\includegraphics[width=2.2in]{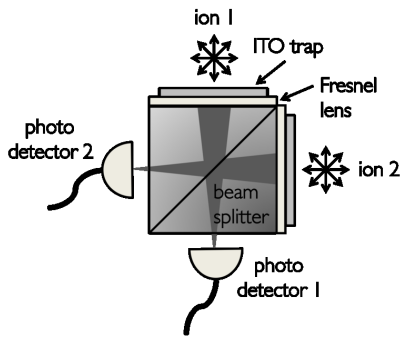}
\caption{\label{fig:entanglement-unit} Diagram of proposed compact entanglement unit (see text).}
\end{figure}

Our measurements establish that ITO is a viable material to use for microfabricated traps, and provide a first demonstration of light collection from ions through a trap with an integrated photodetector. The ability to form transparent traps opens up many possibilities to integrate ions with devices to transfer and detect light efficiently in a scalable architecture. One particularly interesting application is a compact entanglement unit, as shown in Figure \ref{fig:entanglement-unit}. Here, transparent traps mounted on adjacent faces of a beam splitter house two (or more) ions to be entangled. Diffractive optics below the traps overlap the images of the two ions on detectors at the opposite faces of the cube, allowing for heralded entanglement generation between the ions \cite{Luo2009}. For current state-of-the-art experiments, relying on bulk optics and light transmission in fibers, the total coupling efficiency for photons is only $\approx0.004$ \cite{Maunz2009}, resulting in an entanglement generation rate of only $\approx2\times10^{-3}$~s$^{-1}$ (when the experiment is repeated at 100~kHz, assuming a branching ratio of 0.005, and a photodetector quantum efficiency of 15\%). For the proposed compact entanglement unit (neglecting losses due to reflection at interfaces, and absorption in materials other than ITO), the coupling efficiency is $\approx0.45$, resulting in a probability of entanglement generation of $\approx30$~s$^{-1}$, which is $\approx10^3$ times higher than the best rates achieved to date.

This work was supported by the MQCO Program with funding from IARPA, and by the NSF Center for Ultracold Atoms.

\end{document}